\documentclass[pra,showpacs,amsfonts,amsmath,floatfix,onecolumn]{revtex4} 

\usepackage{graphicx}
\usepackage{epstopdf}
\usepackage{amssymb}
\usepackage{amsmath}
\usepackage{xspace}
\usepackage{color}

\begin{document}

\title{Determination of the maximum Global Quantum Discord via measurements of excitations in a cavity QED network}
\author{Raul Coto Cabrera}
\affiliation{Instituto de F\'{i}sica, Facultad de F\'{i}sica, Pontificia Universidad Cat\'{o}lica de Chile, Casilla 306, Santiago, Chile}
\email{rcoto@uc.cl, morszag@fis.puc.cl}
\author{Miguel Orszag}
\affiliation{Instituto de F\'{i}sica, Facultad de F\'{i}sica, Pontificia Universidad Cat\'{o}lica de Chile, Casilla 306, Santiago, Chile}
\email{morszag@fis.puc.cl}


\begin{abstract}
Multipartite Quantum Correlations is one of the most relevant indicator of the quantumness of a system in many body systems. This remarkable feature is in general difficult to characterize and the known definitions are  hard to measure. Besides the efforts dedicated to solve this problem, the question of which is the best approach remains open. In this work, we study the Global Quantum Discord ($GQD$) as a bipartite and multipartite measure. We also check the limits of this definitions and  present an experimental scheme to determine the maximum of the $GQD$ via the measurements of the system`s excitations, during the time evolution of the present system.
\end{abstract}

\maketitle

\section{Introduction}

Quantum correlations has been a hot topic during the last years due to their powerful applications in quantum information and computational tasks \cite{nielsen,caves}. For bipartite states, different measures as \textit{Entanglement}($E$) \cite{wootters1} and \textit{Quantum Discord}($QD$) \cite{QD, QD1} are already well understood. Although, some times for multipartite systems, there are correlations which are not detected by the previous measurements. Many attempts of extending the bipartite correlations to the multipartite case have been made \cite{wootters,zambrini,fanchini,sarandy}, but still questions remain 
about these generalizations.  
One of the first approach was the \textit{Tangle} \cite{wootters}, that is related with $E$, but is difficult to compute for mixed states. Next, in another endeavor, \textit{Global Quantum Discord}($GQD$) was defined in Ref.\cite{sarandy}. This new measurement of correlations is a straight extension from the bipartite to multipartite case, it is symmetric and obeys monogamy properties. These unique advantages suggest the $GQD$ as a resource for quantum information processing.
  
More recently, much attention has been paid to the application of $GQD$ and it's connection with criticality \cite{sarandy,campbell}, as the detection of phase transitions \cite{phase_transition,phase_transition2}. Nevertheless, some questions are open, for example: is it  possible to measure the $GQD$ experimentally, or know when it reaches it's maximum value? To answer this question, we first will study the distribution of excitations in the system, and see how this distribution can affect the $GQD$. We also propose a model, which is a cavity QED system, where the $GQD$ has not been studied yet.

Over the past decades, cavity QED systems have been extensively researched, and several advantages, theoretical and experimentally, are known about these systems \cite{experiment,cavities, victor}. The development of experimental techniques for their manipulation with an unprecedented level of control, as well as performing measurements inside the cavity are desirable features when choosing our model.

This paper is organized as follows: in section II, we describe our system, the Hamiltonian and write a generalized master equation, where the Lindblad terms result from the coupling of each cavity to it's own thermal reservoir at zero temperature. In section III, we give a brief outline of the \textit{Global Quantum Discord}. In section IV, we present the main results of this paper, related to the applicability of $GQD$ and we discuss our ability to gain, experimentally, information about this magnitude. Finally, section V is devoted to the conclusions.

\section{The Model}

\begin{figure}[ht]
\centering
\includegraphics[width=8.3cm]{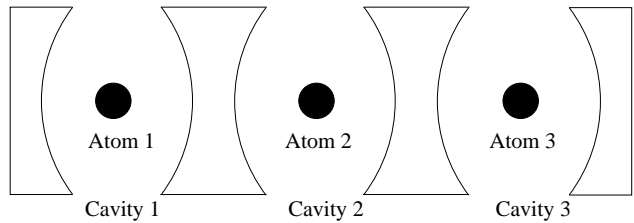}
\caption{Three Coupled Cavity-Atom Systems.}
\label{fig1}
\end{figure}

We have three coupled cavities, as shown in Fig.(\ref{fig1}), where each cavity interacts with a single atom and it's own reservoir. We choose Rydberg atoms with principal quantum numbers $51$ and $50$, where the transition is at $51,1$ GHz. The atom cavity strength coupling($g$), corresponds to an interaction time of $1$ $\mu$s. The photon life time inside the cavity is $T_{cav}=1$ ms \cite{haroche1,haroche2}. The coupling between the cavities($J$) is about $10^{-2}g$. We scale the time in the figures with $\gamma=10^5$.

The Hamiltonian of the system, in the basis of the dressed states(polaritonic) \cite{angelakis}, is given by:

\begin{equation}
\mathit{H}=\sum_{i=1}^3(\omega_{i}-g_{i})|E\rangle_i\langle E|+ +\sum_{i=1}^{2}\frac{J_i}{2}(L_i^{\dagger}L_{i+1} + L_i^{-}L_{i+1}^{\dagger})
\end{equation}

where $|E_i\rangle=\frac{1}{\sqrt{2}}(|1,g\rangle_i-|0,e\rangle_i)$ and $|G_i\rangle=|0,g\rangle_i$ are the dressed states, corresponding to excited and ground state respectively. The other operators $\mathit{L^\dagger_i}=|E_i\rangle\langle G_i|$ and $\mathit{L^-_i}=|G_i\rangle\langle E_i|$ are to create or destroy those states. So we can consider polaritons as  two-level systems. We just can have one photon, at most, because due to photon blockade, double or higher occupancy of the polaritonic states is prohibited \cite{blockade1,blockade2}.\\

The main source of dissipation originates from the leakage of the cavity photons due to imperfect reflectivity of the cavity mirrors. A second source of dissipation, corresponding to atomic spontaneous emission,  will be neglected assuming long atomic lifetimes.

An approach to model the above mentioned losses, in the presence of single mode quantized  cavity field, is using the microscopic master equation, which goes back to the ideas of Davies on how to describe the system-reservoir interactions in a Markovian master equation \cite{davies}. For a three-cavity-system at zero temperature, the master equation is \cite{raul,serafini}:

\begin{equation}\label{em2}
\dot{\rho}(t)=-i\left[\mathit{H_s},\rho(t)\right]+\sum_{n=1}^3\sum_{\omega>0}^{\infty}\gamma_n(\omega)\left( \mathit{A_n}(\omega)\rho(t)\mathit{{A}^\dagger_n}(\omega)-\frac{1}{2}\left\lbrace \mathit{{A}^\dagger_n}(\omega)\mathit{A_n}(\omega),\rho(t) \right\rbrace\right)   
\end{equation}

where $\mathit{A_n}$ correspond to the Davies's operators. The sum on $n$ is over all the dissipation channels and the decay rate $\gamma_n(\omega)$ is the Fourier transform of the correlation functions of the environment \cite{petruccione}.

The $\mathit{A_n}$ operators are calculated as follows:

\begin{equation}\label{operatorA}
\mathit{A_n}(\omega_{\alpha\beta})=|\phi\rangle_{\alpha}\langle\phi|a_n|\phi\rangle_{\beta}\langle\phi|
\end{equation}


\section{Global Quantum Discord}

In the original proposal \cite{QD}, $QD$ was defined as a mismatch between quantum analogs of classically equivalent expressions of the mutual information.

\begin{equation}\label{QD}
\mathit{QD}(\rho_{AB})=\mathit{I}(\rho_{AB})-\mathit{J}(\rho_{AB})
\end{equation}

The mutual information  $\mathit{I}(\rho_{AB})$ of two subsystem can be expressed as 

\begin{equation}
\mathit{I}(\rho_{AB})=S(\rho_A)-S(\rho_A\vert\rho_B),
\end{equation}

where $S(\rho)=-Tr(\rho\log_2\rho)$ is the von Neumann entropy, and $S(\rho_A\vert\rho_B)=S(\rho_{AB})-S(\rho_B)$.

The classical correlation $\mathit{J}(\rho_{AB})$ is defined as the maximum information that one can obtain from $A$ by performing a measurement on $B$, and in general this definition is not symmetric:

\begin{equation}\label{CC}
\mathit{J}(\rho_{AB})=\max_{\lbrace \Pi_B^k \rbrace}[S(\rho_A)-S(\rho_{AB}|\lbrace \Pi_B^k\rbrace)],
\end{equation}    

where $\lbrace \Pi_B^k\rbrace$ is a complete set of projectors performed on subsystem $B$ and $S(\rho_{AB}|\lbrace \Pi_B^k\rbrace)=\sum_k p_k S(\rho_A^k)$. The reduced density operator $\rho^k$ associated with the measurement result $k$ is:

\begin{equation}
\rho^k=\dfrac{1}{p_k}(\mathit{I}\otimes \Pi_B^k)\rho(\mathit{I}\otimes \Pi_B^k)
\end{equation}

with $\mathit{I}$ the identity operator.

Notice that $\mathit{I}(\rho_{AB})$ can be rewritten in terms of the relative entropy, $S(\rho\Vert\sigma)=Tr(\rho\log_2\rho-\rho\log_2\sigma)$, as:

\begin{equation}
\mathit{I}(\rho_{AB})=S(\rho_{AB}\Vert\rho_A\otimes\rho_B)
\end{equation}

Also, by symmetrizing the definition through the introduction of bilateral measurements, and after some algebra we get a new definition of $QD$, given by:

\begin{equation}\label{bgqd}
GQD(\rho_{AB})= \min_{\lbrace\Pi_A^j\otimes\Pi_B^k\rbrace}[S(\rho_{AB}\Vert\Phi_{AB}(\rho_{AB}))-S(\rho_A\Vert\Phi_A(\rho_A))-S(\rho_B\Vert\Phi_B(\rho_B))]
\end{equation}

with $\Phi(\rho_{AB})=\sum_{j,k}(\Pi_A^j\otimes\Pi_B^k)\rho_{AB}(\Pi_A^j\otimes\Pi_B^k)$. From Eq.(\ref{bgqd}) the generalization to multipartite discord is evident,

\begin{equation}\label{GQD}
GQD(\rho_{A_1 ... A_N}) =\min_{\lbrace\Pi^k\rbrace}[S(\rho_{A_1...A_N}\Vert\Phi(\rho_{A_1...A_N}))-\sum_{j=1}^{N} S(\rho_{A_j} \Vert\Phi(\rho_{A_j}))]
\end{equation}

where $\Phi(\rho_{A_j})=\sum_{k}\Pi_{A_j}^k\rho_{A_j}\Pi_{A_j}^k$ and $\Phi(\rho_{A_{1} ... A_{N}})=\sum_k\Pi_k\rho_{A_1 ... A_N}\Pi_k$, with $\Pi_k=\Pi_{A_1}^{k_1}\otimes \dots \otimes\Pi_{A_N}^{k_N}$ and $k$ denoting the index string $(j_{1} ... j_{N})$.

\section{Results}

\subsection*{Genuine Tripartite Measure}

\begin{figure}[ht]
\centering
\includegraphics[height=5cm,width=8cm]{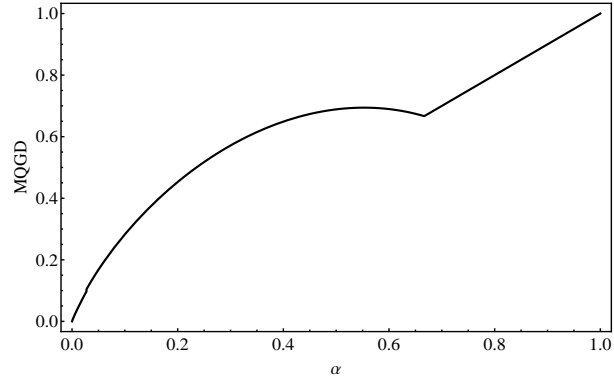}
\caption{Genuine tripartite global quantum discord increases from zero to one when the initial state goes from Bell to $GHZ$ state. }
\label{fig2}
\end{figure}

It has been shown that $GQD$ is a multipartite measurement \cite{sarandy,campbell}, that not only measures tripartite quantum correlations, as the \textit{Tangle} defined by Wootters \cite{wootters}, but also bipartite correlations. This statement can be illustrated with the following example.
 If we prepare our system initially in a mixture of a genuine tripartite correlated state(GHZ) and a bipartite Bell state,

\begin{equation}\label{initial}
\rho(0)=\frac{\alpha}{2}(|EEE\rangle\langle EEE|+|GGG\rangle\langle GGG|)+\frac{(1-\alpha)}{2}(|\Psi\rangle\langle\Psi|\otimes |G\rangle_2\langle G|)
\end{equation}

with $|\Psi\rangle=(|E_1G_3\rangle+|G_1E_3\rangle)$, as $\alpha$ increases from zero to one, the system goes from bipartite to tripartite correlations, but $GQD=1$ for all $\alpha$. The question is, what happens when we eliminate all the bipartite quantum discord?  In Fig.(\ref{fig2}) we plot the function $MGQD=GQD_{123}-GQD_{12}-GQD_{13}-GQD_{23}$, for the same initial state in Eq.(\ref{initial}). Notice that for $\alpha=0$ there is no multipartite correlation and for $\alpha=1$ the $MGQD$ is one, as expected from a $GHZ$ state. Near to $\alpha=0.7$ the function has a point where the derivative does not exist, this is because of the change  in the angles during the numerical minimization.  

\begin{figure}[ht]
\centering
\includegraphics[height=5cm,width=8cm]{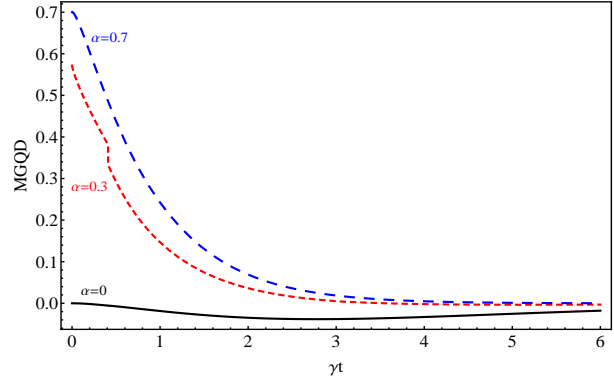}
\caption{For $\alpha=0$ the $MGQD$ becomes negative, indicating that the $GQD$ does not include separately bipartite and tripartite correlations. $T_{cav}=10$ $\mu$s.}
\label{fig3}
\end{figure}

At this point, it seems that there is no problem with the new definition of genuine multipartite correlation. However, when we check the time evolution for $MGQD$, particularly for $\alpha=0$, the function becomes negative at certain times, see Fig.(\ref{fig3}). We also tried  the Werner's state, obtaining similar results. This negative behavior of $MGQD$ is enhanced when the initial condition is near a pure bipartite correlated state. 

A first approach to solve this problem can be the use of the monogamy restrictions \cite{monogamy1,monogamy2}, where the exact solution is lost, but at least we can estimate a upper bound for the genuine tripartite correlations. From references \cite{monogamy1} and \cite{monogamy2}, we write two monogamy relations:

\begin{equation}\label{DR1}
GQD(A:B:C)\geq GQD(A:B) +GQD(A:C)
\end{equation}

\begin{equation}\label{DR2}
GQD(A:B:C)\geq GQD(A:B) +GQD(B:C)
\end{equation}

The authors of these two papers define a ``Residual $GQD$"($D_R$) as the difference between the left hand and right hand side of above equations. The problem with the definition of $D_R$ is that is non symmetric with respect to the pairwise combinations. Instead, we define a new $D_R$, based on the above equations, getting:

\begin{equation}\label{DR3}
GQD(A:B:C)\geq \frac{2}{3}(GQD(A:B)+GQD(B:C)+GQD(A:C))
\end{equation} 

\begin{figure}[ht]
\centering
\includegraphics[height=5cm,width=8cm]{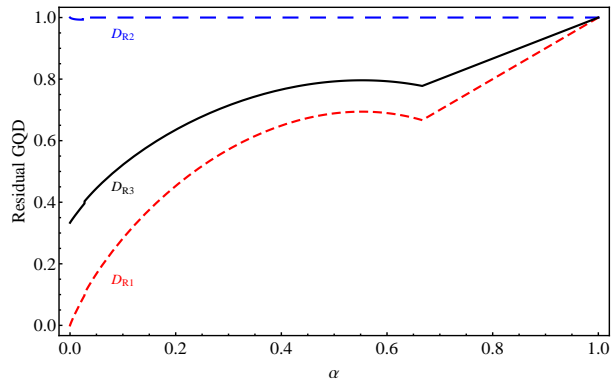}
\caption{ Residual $GQD$ corresponding to our definition($D_{R3}$) represents better the monogamy restriction, since it is a good approximation independently of the initial condition.}
\label{fig4}
\end{figure}

In Fig.(\ref{fig4}) we reported the comparison between $D_{R1}$, $D_{R2}$ and $D_{R3}$ from equations (\ref{DR1}),(\ref{DR2}) and (\ref{DR3}) respectively. Already from the initial state there are differences among the three curves. Notice that the residual global discord corresponding to Eq.(\ref{DR1})(red-dotted), seems to be the most restrictive one. Nevertheless, that can be easily changed by starting with a bipartite correlation of cavities $2$ and $3$, instead cavities $1$ and $3$, which will change $D_{R2}$ to be the most restrictive one. But, our approach remains very well independent of the initial condition, as it includes all possible combination of pairwise correlations. 

\begin{figure}[ht]
\centering
\includegraphics[height=5cm,width=8cm]{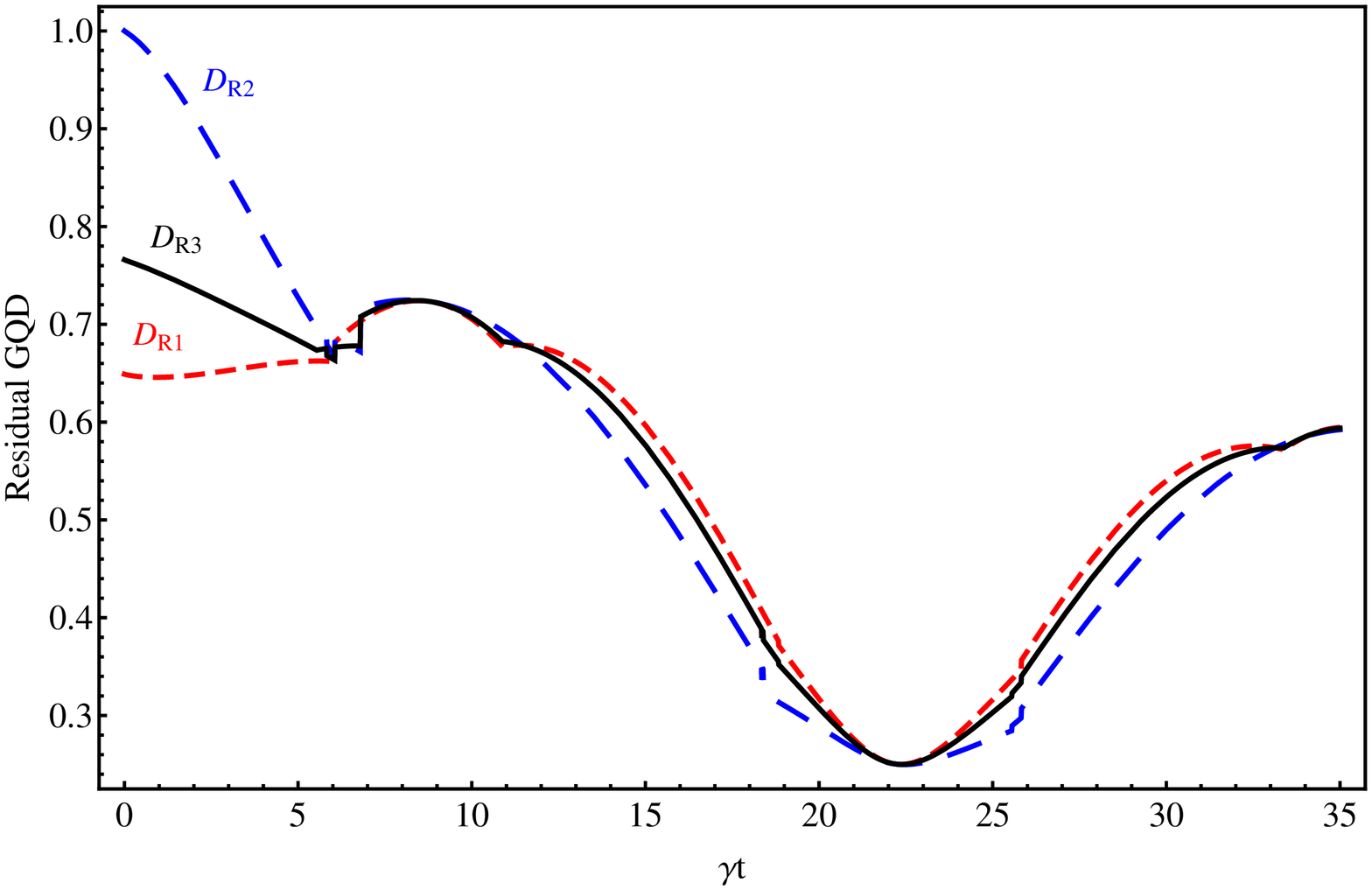}
\caption{All definitions are close, during the time evolution of the system. We observe that $D_{R3}$ remains between the other two, again showing more stability to variations of the initial conditions.}
\label{fig5}
\end{figure}

Next, we analyzed the time evolution of the above definitions for $\alpha=0.4$. In Fig.(\ref{fig5}) we show that certainly $D_{R1}$ and our definition $D_{R3}$ are close. However, $D_{R1}$ is highly sensitive to initial conditions, which is not the case of $D_{R3}$, so we conclude that  $D_{R3}$ is more suitable to describe the quantum correlations, for any initial condition. 

\subsection*{Estimation of the $GQD$ by means of the excitation probabilitities of the subsystems  }

Quantum Correlation measurements are very important for quantum information and quantum computation, and even now is difficult to do it \cite{davidovich}, especially for higher correlations as the tripartite one. However, there is a connection between the localization of the excitations throughout the system and the quantum correlations of it's parts. To illustrate this, we first consider a typical bipartite Bell state $|\phi\rangle=\frac{1}{\sqrt{2}}(|10\rangle+|01\rangle)$. It is well known that this state is maximally correlated, but we also notice that the probability of finding an excitation in each subsystems is $1/2$. In other words, we could say in this example, that when the subsystems are highly correlated, the excitations are equally distributed through them.

\begin{figure}[ht]
\centering
\includegraphics[height=5cm,width=8cm]{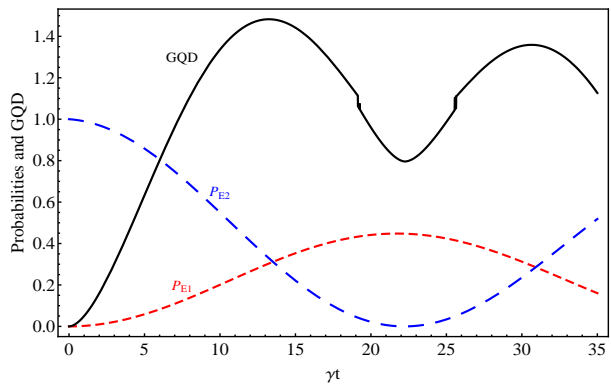}
\caption{$GQD$ reaches it's maximum when the three probabilities cross at certain time. Due to symmetry, $P_{E3}=P_{E1}$. }
\label{fig6}
\end{figure}

In our system, things are more complicated, since we have three cavities and we could have up to three excitations. Nevertheless, the same rule applies. For example, let us assume that initially we have one excitation in cavity $2$, and let $P_{E1}$, $P_{E2}$ and $P_{E3}$ be the probabilities of finding the polariton in cavities $1$, $2$ and $3$ respectively. In Fig.(\ref{fig6}), we plot the time evolution of $GQD$ and these three probabilities. We can readily see that when the three probabilities cross at a certain time, the $GQD$ reaches it's maximum value, as in the case of two qubits. Thus we believe that the $GQD$ is  associated with disorder or equal distribution of the excitations among the three cavities.

\begin{figure}[ht]
\centering
\includegraphics[height=5cm,width=8cm]{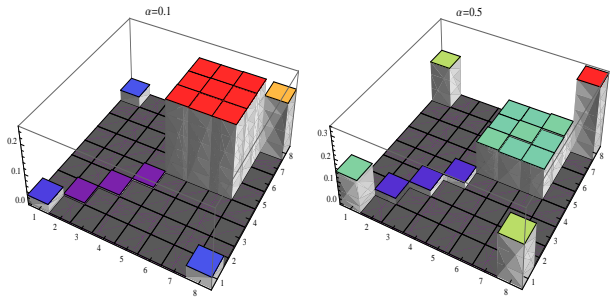}
\caption{Density operator's elements for the initial state in Eq(\ref{initial}). The maximum of $GQD$ is reached when $P_{E1}=P_{E2}=P_{E3}$ and the off-diagonal components do not vanish.} 
\label{fig7}
\end{figure}

Similar results are also observed for the state in Eq.(\ref{initial}), see Fig.(\ref{fig7}). Here we show  the matrix's elements of the density operator for $\alpha=0.1$ and $\alpha=0.5$. We used the standard basis: $\vert 1\rangle =\vert EEE \rangle$,$ \vert 2\rangle =\vert EEG\rangle$,$\vert 3\rangle =\vert EGE\rangle$,$\vert 4\rangle =\vert GEE\rangle$,$\vert 5\rangle =\vert EGG\rangle$,$\vert 6\rangle =\vert GEG\rangle$,$\vert 7\rangle =\vert GGE\rangle$,$\vert 8\rangle =\vert GGG\rangle$. Each graphic corresponds to the maximum of the $GQD$. Notice that again the three probabilities, associated to the $\vert5\rangle\langle5\vert$,$\vert6\rangle\langle6\vert$ and $\vert7\rangle\langle7\vert$ matrix elements, are equal.

The presence of the quantum correlations is related to the off diagonal elements of the density matrix. For long times, these elements as well as the correlations tend to disappear due to the losses. 

As we saw, one of the advantages of the $GQD$ is that for any mixed initial bipartite and tripartite state, with only one measurement we can estimate how correlated the subsystems are.
Then, one could experimentally detect when the maximum $GQD$ is reached by measuring the polaritons in the cavities \cite{measure_polariton}.
To summarize,  the $GQD$ can provides us valuable information about any class of multipartite correlations and furthermore, this can be experimentally observed by measuring the excitations of our system.

\section{Summary and Conclusions}

We analized the Global Quantum Discord, as a measure of the joint bipartite and tripartite correlations.
We showed it´s limitations to detect a genuine tripartite correlation, since negative values show up.
However, we presented an upper bound which turned out to be a good estimation, valid for any initial condition. Then we studied the relation between the disorder of the system and the $GQD$.
Our goal was to associate the $GQD$ with some experimentally measurable quantity, such as the degree of excitation of each sub-system.
We found that when excitations were nearly equally distributed, among the various sub-systems, the $GQD$ reached it´s maximal value.

Moreover, the sensitivity of this measure, which is certainly related with the bilateral projection and the minimization process, seems very interesting for it's different applications.  In order to illustrate this feature, we focus on the sudden transition effect \cite{sudden_trans,sudden_trans2}. This effect depends strongly on the the initial conditions, and it can be seen only when some restrictions are fulfilled. For example, we start with the initial state proposed in reference \cite{sudden_trans} for the cavities $1$ and $3$, and assume for the second cavity to be in an excited or ground state.

\begin{equation}
\rho(0)=\begin{pmatrix}
(1+c_3) &0 &0 &(c_1-c_2) \\ 0 &(1-c_3)  &(c_1+c_2) &0 \\ 0 &(c_1+c_2) &(1-c_3) &0\\ (c_1-c_2) &0 &0 &(1+c_3)
\end{pmatrix}\otimes|\mathit{i}\rangle_2\langle\mathit{i}|
\end{equation}

where the matrix is in the basis:$|EE\rangle$,$|EG\rangle$,$|GE\rangle$,$|GG\rangle$, and $|\mathit{i}\rangle=\left\lbrace|E\rangle,|G\rangle\right\rbrace$.

\begin{figure}[ht]
\centering
\includegraphics[height=5cm,width=8cm]{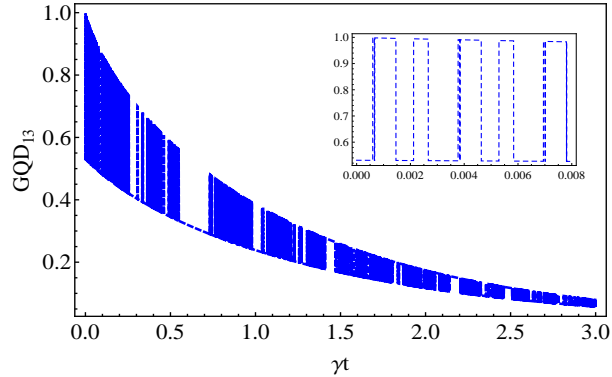}
\caption{Sudden changes in the bipartite global quantum discord for the cavities $1,3$. $T_{cav}=10$ $\mu$s.}
\label{fig8}
\end{figure}

In Fig.(\ref{fig8}) we plotted the quantum discord, defined in Eq.(\ref{bgqd}) between cavities $1$ and $3$, when cavity $2$ is initially in the state $|G\rangle$, and weakly coupled to the other two cavities. The parameters are: $c_1=1$,$c_2=-c_3$ and $c_3=0.8$. The inset corresponds to a zoom at the beginning of the curve.
 We observe rapid oscillations that have not been reported before, for this particular measure, and also abrupt changes in the derivative, which is  quite unusual.
  We did the same for $QD$ defined in Eq.(\ref{QD}), following two different approaches  \cite{alber,caldeira}, and we did not find such effects in neither case. This evidences that $GQD$, proposed in reference \cite{sarandy} is more sensitive than the others.

\section{Acknowledgements}

M. Orszag acknowledges financial support from Fondecyt, Project $1100039$ and Programa de Investigacion asociativa anillo ACT-1112. R. Coto thanks the support from the Pontificia Universidad Cat\'olica de Chile.

\end{document}